# Meta-Generalization for Multiparty Privacy Learning to Identify Anomaly Multimedia Traffic in Graynet

Satoshi Nato* and Yiqiang Sheng

*Abstract*—Identifying anomaly multimedia traffic in cyberspace is a big challenge in distributed service systems, multiple generation networks and future internet of everything. This letter explores meta-generalization for a multiparty privacy learning model in graynet to improve the performance of anomaly multimedia traffic identification. The multiparty privacy learning model in graynet is a globally shared model that is partitioned, distributed and trained by exchanging multiparty parameters updates with preserving private data. The meta-generalization refers to discovering the inherent attributes of a learning model to reduce its generalization error. In experiments, three meta-generalization principles are tested as follows. The generalization error of the multiparty privacy learning model in graynet is reduced by changing the dimension of byte-level imbedding. Following that, the error is reduced by adapting the depth for extracting packet-level features. Finally, the error is reduced by adjusting the size of support set for preprocessing traffic-level data. Experimental results demonstrate that the proposal outperforms the state-of-the-art learning models for identifying anomaly multimedia traffic.

## I. INTRODUCTION

Graynet refers to a networking system with unexpected components that may include darknets or untrusted networks [1]. The darknet is a network that is configured for collecting suspicious data using botnets to trap adversaries, distributes illegal material using dark nodes, or operates without identities using private nodes. Anomaly traffic and malicious node identifications [2] are typical challenges for distributed service systems with feature learning for rich media [3], cross-media [4] or multimedia [5]. In retrospect, the first generation computer network is a simple media graynet oriented to terminals with no local resources and applications using the client-server model connected by modems and public telephone networks. The second generation computer network is a homogenous media graynet that provides communication among hosts with locally sharable resources and applications for users connected via interface message processors. The third generation computer network is a heterogeneous media graynet based on international standardizations including Internet protocol suite and open systems interconnection reference model to provide end-to-end transmission by focusing on the interoperability of communication systems for multimedia service.

Furthermore, the next generation networks (NGNs) can be partially regarded as a graynet that focuses on 5W personal communication (i.e. whoever, whenever, wherever, whomever and whatever) and secure relationship between network and human, which can be the network on human, the network in human, or the network with human. Nowadays, with the research and development of 5G mobile communications, network function virtualization, software defined network and information centric networking, the testbeds for NGNs include global environment for network innovations (GENI) [6] and PlanetLab, the project of future Internet design (FIND) [7] with user controllable routing, the project of challenge one with autonomic Internet in the seventh framework program to provide the capabilities of scalable virtualized network service and sensible optimized resource management, and the project from the national institute of information and communications technology in Japan to support secure data accesses and scalable network functionalities. It results in a great issue of multimedia service cybersecurity.

In future, it is increasingly important to make communication smart and safe for internet of everything that brings diverse entities including human, things, environmental data, information, knowledge and context-aware processes connected as a pervasive cyberspace. Furthermore, the upcoming distributed service system security should adapt specific domains with more advanced applications such as virtual reality, augmented reality, mixed reality, holography, autonomous vehicles, remote surgeries, and intelligent robots as human-machine interfaces. Many specific domains require inherent security with increasingly high performance such as high bandwidth, massive connections, on-demand deterministic latency, lossless transmission, embedded credibility, traceability and reliability. The internet with intelligence such as robots tends to be connected as socialized citizens in a human-machine convergent society. It leads to a big challenge for multimedia service security and privacy.

The rest of this letter is organized as follows. Section II covers the motivation. Section III is the contribution. Section IV presents the proposal. Section V introduces the evaluation. Section VI discusses the results. Section VII concludes the letter.

## II. MOTIVATION

The research is motivated by the hardness of identifying anomaly traffic in a graynet to emphasize the necessity of multimedia service security and privacy with the federation of a large number of clients based on private data. In the ever-changing cyberspace with rapidly evolving attackers and defenders, it is increasingly crucial to ensure both security and privacy. The traditional methods based on port, payload and statistics fail to detect unexpected attacks. The attacks can be divided into multiple categories such as brute force attack, distributed denial of service (DDoS) attacks, botnet attack, web attack, infiltration attack, remote to local attacks, user to root attack, and probing attacks. Signature-based pattern recognition tools such as Suricata and Snort [8] were considered as a part of intrusion detection. The relevant feature selection and pattern matching methods such as Rabin-Karp [9] and q-gram [10] algorithm are used to compare with known attack patterns called rules in a firewall and antivirus software. However, the above mentioned methods require massive manual data preprocessing or manual feature extraction.

---

* satoshint@yeah.com <span>I am actually dying due to widely metastasized cancer since 2019, though it is a pity there are still so many unpublished researches so far.</span>

The progress of artificial intelligence has enabled researchers to make breakthroughs in anomaly multimedia traffic classification for distributed service system by using deep learning models [11] such as deep logic network [12], long short-term memory [13], and attention with generative adversarial network [14] to automatically learn patterns and create statistic rules. In fact, many researches demonstrate that deep learning models reach better performance than traditional models, but the generalization error is still a big issue to identify and classify anomaly multimedia traffics.

## III. Contribution

The primary contribution is to explore the inherent attributes of generalization for a multiparty privacy learning model in graynet to improve the performance of anomaly multimedia traffic identification. The model is partitioned, distributed and trained to optimize its shared parameters by exchanging multiparty updates with preserving private data. Three principles of meta-generalization are explored and tested as a defense-in-depth strategy. The multiparty privacy learning model is utilized to keep an on-demand balance between generalization and personalization for anomaly multimedia traffic identification. It differs from a firewall in that it monitors the incoming and outgoing traffic defined in a set of rules to raise alarms for detecting a threat. It can be summarized as follows. First, the performance of anomaly multimedia traffic identification is improved by changing the dimension of byte-level imbedding space. Next, it is improved by adapting the network depth of packet-level feature extraction to get relevant features. Finally, the performance is further improved by adjusting the size of support set of traffic-level preprocessing to make use of contextual information.

## IV. Proposal

Multiparty privacy learning refers to a shared multi-dimensional deep network model structured by a support set that aggregates many private updates by multiparty parametric exchanges among clients, brokers and servers. The model exchanges private parameters by preserving private data. In case of the application of anomaly multimedia traffic identification, the input is a split traffic and its attributes, while the output is a known label to correctly identifying an anomaly multimedia traffic. The split traffic is sampled from public datasets, while the attributes of the split traffic are sampled from private datasets.

A graph is designed to describe the action and relation in graynet as follows. The graph connects subjective, objective and hidden vertices including darknets and brokers. The clients refer to subjective vertices, and the servers are objective vertices. The brokers and darknets are denoted by hidden vertices. A granular entity refers to a cluster of clients and servers interspersed with darknets and brokers.

Let $G$ $(S, O, H, A, R)$ denote the above mentioned action-relational graph for coalition formation with granular entities. $D_a$ denotes actional dataset, while $D_r$ denotes relational dataset. $S=\{s_l(b, f)\}$ is a set of subjects $s_l$ such as clients with behaviors and feedbacks $(b, f)$. $O=\{o_m(s, p)\}$ is a set of objects $o_m$ such as servers with states and probabilities $(s, p)$. $H=\{h_{ue}\}$ is a set of hidden vertices $h_{ue}$ such as darknets or brokers. $A=\{a_n(t_s, t_e)\}$ is a set of actions $a_n$ between vertices with starting time $t_s$ and ending time $t_e$. $R = R_s + R_o = \{r_{us}(w_{s+}, w_{s-})\} + \{r_{uo}(w_{o+}, w_{o-})\}$ is a set of two classes of relations between subjects $r_{us}$ with bidirectional weights $(w_{s+}, w_{s-})$ and between objects $r_{uo}$ with bidirectional weights $(w_{o+}, w_{o-})$. The above mentioned $i, l, m, n, u_e, u_s, u_o, v, h$, and $j$ are integers.

$G_N$ $(S, O, A, R)$ denotes a graynet that granularizes darknets based on a granularity $t_g$ to form the coalition of multiparty computation. The granularity $t_g$ can be user-defined or calculated by an experimental formula, which may depend on computational cost, communication cost, multiparty data size, multiparty parameter size, fitting factor for early stop, runtime prediction, partitioning factor, and maximum value of latency divided by bandwidth.

The basic model of multiparty privacy learning consists of a globally shared neural net, named congruity net, on a broker, many partitioned parts of the congruity net on servers, a large number of neural nets on clients, named federated nets. The congruity net is tuned by exchanging parameters of many federated nets. Let $A^*=\{a_{n*}\}$ be a set of congruity actions, and let $R^*=\{r_{v*}\}$ denote a set of congruity relations. $Y(S^s)$ presents the output of congruity net based on a support set $S^s=\text{supp}_k(Y)$. $Y(S^{s1})$ is the output based on a complementary support set $S^{s1}$, and $Y(S^{s2})$ denotes the output based on another complementary support set $S^{s2}$ with a soft coefficient $\beta$. A set of the granular subjective entities is marked as $S^*$, while a set of the granular objective entities is marked as $O^*$.

The set of federated net parameters is marked as $\Theta_s=\{\theta_{s1}, \theta_{s2}, ..., \theta_{sN}\}$, $N$ is the number of federated net parameters. The federated domain is marked as $S_D=\{X_s, P(x_s)\}$, where $X_s$ is a feature space, $x_s=\{x_{si}\}$ belongs to $X_s$, and $P(x_s)$ is the probability distribution with respect to $x_s$ in the edge of space $X_s$. The federated net is marked as $S_T=\{Y_s, P(Y_s/X_s)\}$, where $Y_s$ is a labelling space, $P(Y_s/X_s)$ is a predictive probability function obtained by training based on samples from the feature space $X_s$ in the domain $S_D$.

The set of congruity net parameters is marked as $\Theta_t=\{\theta_{t1}, \theta_{t2}, ..., \theta_{tM}\}$, $M$ is the number of congruity net parameters. The congruity domain is marked as $T_D=\{X_t, P(x_t)\}$, where $X_t$ is a feature space, $x_t=\{x_{ti}\}$ belongs to $X_t$, and $P(x_t)$ is the probability distribution with respect to $x_t$ in the edge of space $X_t$. The congruity net is marked as $T_T=\{Y_t, P(Y_t/X_t)\}$, where $Y_t$ is a labelling space, $P(Y_t/X_t)$ is a predictive probability function obtained by training based on samples from the feature space $X_t$ in the domain $T_D$.

Let $\Theta_h = \{(\theta_{h1}, \theta_{h2}, ..., \theta_{hj}, ..., \theta_{hv})\}$ be a set of layer-wise parameters, where $h$ is an order number of layers with $0 \leq h \leq H_{max}$ +1, $H_{max}$ is the adaptable number of hidden layers, $h = 0$ means the input layer, $h = H_{max}$+1 means the output layer, $v$ is the number of parameters in a given $h$-th layer, and each $\theta_{hj}$ includes a connection vector $C_j = (c_{1j}, c_{2j}, ...,)$ and a bias scalar $b_j$.

Let $D_g$ denote public dataset. $\{D_{ps}\}$ is a set of private datasets. $E(\Theta)$ is a user-defined error function, $\Theta^*$ is an optimized set of parameters, and $E^* = E(\Theta^*)$ is the generalization error. $H_{fl}$ is the size of support set for preprocessing traffics. $H_{pl}$ is the depth of packet-level feature extraction. $H_{bl}$ is the dimension of byte-level imbedding. Following that, a multiparty privacy learning algorithm in a graynet is designed to improve the performance of anomaly multimedia traffic identification.

---

**Algorithm:** Multiparty privacy learning

**Input:**

---

* satoshint@yeah.com 

$D_g$: public dataset in servers
$\{D_{ps}\}$: a set of private datasets in clients
$D_a$: actional dataset between clients and servers
$D_r$: relational dataset between clients or between servers

**Output:**

$E^*$: generalization error

**Hyperparameters:**

$H_{fl}$: size of support set for traffic-level data preprocessing
$H_{pl}$: depth of packet-level feature extraction
$H_{bl}$: dimension of byte-level imbedding

**Parameters:**

$\Theta$: initial parameters
$\Theta^*$: optimized parameters
$G\ (S, O, H, A, R)$: action-relational graph
$G_N\ (S, O, A, R)$: graynet that granularizes darknets
$t_g$: granularity
$b_s$: identity of each server
$b_c$: identity of each client

**Broker ($E^*$):**

Create a globally shared model with initial parameters $\Theta = \{\theta_j\}$
Partition the model and $D_g$ according to granularity
Assign learning tasks to servers according to partitioning
For each assigned server $b_s$ in $H$ in parallel
    Execute the Server ($b_s$, $\Theta^*$, $G_N$)
    Update $\Theta^*$ using validation data in $D_g$
Update $G_N$ and partitioning
Place learning jobs to clients according to $G_N$ and partitioning
For each placed client $b_c$ in $S^*$ in parallel
    Execute the Client ($b_c$, $\Theta^*$, $G_N$)
    Update $G_N$ and $\Theta^*$
Get $E^*$ with $G_N$ and $\Theta^*$ using the selected test data with labels
Return $E^*$ as output

**Server ($b_s$, $\Theta^*$, $G_N$):**

For $h$ from 0 to $H_h$+1
    For all batches of data in $D_g$
        Optimize $E_h\ (\Theta)$ using error back propagation
    If $h = H_{max}$ +1
        Training the model until saturation to get $E_{Hh+1}$
        If $E_{Hh+1} < E_{Hmax}$ and $H_{max} < H_{pl}$
            $H_{max} = H_{max} + 1$
        End if
    End if
Choose at least one seed from $S$ and $O$
For all chosen seeds
    Cluster hidden vertices into the seeds of objects to get $O^*$
    Cluster hidden vertices into the seeds of subjects to get $S^*$
For all batches in $D_a$
    Optimize $E\ (\Theta)$ to get $\Theta^{a*}$ by error back propagation, subject to $Y(S^{c1}) = 0$, where $S^{c1} = \text{supp}^c_{\ k}(Y)$
For all batches in $D_r$
    Optimize $E\ (\Theta^{r*})$ to get $\Theta^*$ by error back propagation, subject to $Y(S^{c2}) = 0$, where $S^{c2} = \text{supp}^c_{\ \beta1k}(Y)$
For all batches in $D_g$
    Optimize $E(\Theta^{r*})$ to get $\Theta^*$ by error back propagation, subject to $Y(S^c) = 0$, where $S^c = \text{supp}^c_{\ \beta2k}(Y)$
Update $G_N$ according to granularity $t_g$
Return $\Theta^*$ and $G_N$

**Client ($b_c$, $\Theta^*$, $G_N$):**

Prepare mini-batches of training data in $D_{ps}$
Analyze traffics in the mini-batches
Add sparse layers with an adjustable size of support set $H_{fl}$

Segment traffic into packets
Add dense layers with an adaptable depth $H_{pl}$
Extract packet-level feature
Normalize each feature to get a vector
Split packet into bytes
Imbed bytes with a changeable dimension $H_{bl}$
For all mini-batches in $D_{ps}$
  Optimize $E(\Theta)$ to get $\Theta^*$
Update $G_N$ according to granularity $t_g$
Return $\Theta^*$ and $G_N$

## V. Evaluation

For a fair comparison in a typical graynet using real infrastructure, all models were tested by unexpected multimedia traffic data in distributed clients with private data combined with three datasets of ISCXTor2016 [15], CICDDoS2019 [16] and MAWI2019 [17]. Tor is a darknet. The ISCXTor2016 is generated by multiple categories of real-world communication tasks based on Tor traffic such as HTTPS browsing, FTP over SSL, Skype chatting, IMAP/SSL email, Bittorrent P2P, YouTube video streaming, VoIP, and Spotify audio streaming with benign nonTor traffic from VPN captured by wireshark and tcpdump. The CICDDoS2019 is generated by collecting the up-to-date DDoS attacks that aim at exhausting the available resource of target machines or the bandwidth of target networks with malicious traffic and providing the feature sets for a taxonomy of weighted DDoS attacks including exploitative and reflective attacks. The captured traffic data of DDoS attacks in pcap files include PortMap, PortScan, NetBIOS, UDP, UDP-Lag, TFTP, SYN, MSSQL, NTP, DNS, LDAP, SNMP, WebDDoS, and SSDP with multiple features such as Timestamp, Traffic Duration, SYN Flag Count, Packet Length Min, and Packet Length Max. The MAWI2019 is publicly available on packet traces, which contain the traffic captured for 15 minutes since 2001 until nowadays, from measurement and analysis on the WIDE Internet archive on a transpacific link between Japan and the USA.

To evaluate the generalization error ($G_E$) for identifying anomaly multimedia traffic, a metric is defined by a balance percentage of incorrectly identified unexpected traffics over the total number of traffics and correctly identified traffics based on confusion matrix.

$$G_E = \frac{\xi FN + (1-\xi)FP}{1 + \xi TN + (1-\xi)TP}.$$

where $\xi$ is a balance coefficient between 0 and 1, the true positive ($TP$) indicates a traffic that is correctly identified to be anomaly (and is abnormal in fact), the true negative ($TN$) means a traffic that is correctly identified to be normal (and is normal in fact), the false positive ($FP$) is a traffic that is incorrectly identified to be normal (but is abnormal in fact), and the false negative ($FN$) is a traffic that is incorrectly identified to be anomaly (but is normal in fact). Besides, the number 1 is added in the formula due to $TP+FN=1$ and $TN+FP=1$.

The experiment was implemented with 10-fold cross-validation for all tested models in a distributed service system with mobile devices for learning the models, which in turn improve the user experience on the devices. It is important for a representative part of the botnets for distributed service to study underlying statistical models, emulations, experiments and their repeatability for validation in case of anomaly multimedia traffic identification.

The real infrastructure includes at least one broker, servers, clients, darknets, data collectors, routers, switches, gateways, access points, and mobile devices. The states of infrastructure are measured by the data collectors interacted among objects, subjects, and hidden vertices. The data are converged to the broker and the servers as the inputs and subsequent feedbacks. All servers are assisted by Google Colab [18] with NVIDIA GPU Tesla K80 24GB GDDR5 and 8 processors of 1.90GHz Intel Core i5-4300U CPU 8GB RAM. The broker is with NVIDIA GPU Grid K2 128GB GDDR5 and 24 processors of 2.10GHz Intel Xeon E5-2620 CPU 32GB RAM. All clients are with NVIDIA GEFORCE GTX 760 GPU 2GB GDDR5 and two processors of 2.50GHz Intel Core i7-4710MQ CPU 16GB RAM.

Without loss of generality, the letter assumes all parallel updates for parameters are synchronous in an environment of bandwidth-limited communication between at least one broker, multiparty servers and clients with a fixed number. Each client is independent with a partitioned model and a local dataset. The selection of clients is uniformly random with a fixed probability. Only one selected client would participates a typical process of optimization for the parameters adjustment of the globally shared model. Each selected client executes local training and exchanges the updated parameters with its server. For the structure of neural networks, the depth is set to be 3 to 13, and the size of support set is set to be from 100 to 500. Each hidden layer is set to be 512 units, and the embedding dimension is adjusted from 16 to 512. The balance coefficient $\xi$ is set to be 0.35 based on many trial experiments.

As shown in Table 1, the generalization errors of the proposal are improved by maximum 1.707%, average 0.159% and minimum 0.003% in comparison to the latest models with respect to the embedding dimension from 16 to 512. As shown in Table 2, the generalization errors of the proposal are improved by maximum 2.804%, average 0.417% and minimum 0.005% in comparison to the latest models with respect to the extraction depth from 3 to 13. As shown in Table 3, the generalization errors of the proposal are improved by maximum 3.616%, average 0.344% and minimum 0.009% in comparison to the latest models with respect to the size of support set from 100 to 500.

**Table 1** The comparison with respect to the embedding dimension.

| Dimension | DLN [12] | LSTM [13] | Attention [14] | Proposal |
|---|---|---|---|---|
| 16 | 1.769% | 0.159% | 0.232% | 0.063% |
| 32 | 0.467% | 0.053% | 0.072% | 0.042% |
| 64 | 0.082% | 0.023% | 0.049% | 0.021% |
| 128 | 0.028% | 0.032% | 0.028% | 0.009% |
| 256 | 0.063% | 0.041% | 0.020% | 0.010% |

* satoshint@yeah.com

I am actually dying due to widely metastasized cancer since 2019, though it is a pity there are still so many unpublished researches so far.

| 512 | 0.088% | 0.098% | 0.018% | 0.012% |

**Table 2** The comparison with respect to the extraction depth

| Depth | DLN [12] | LSTM [13] | Attention [14] | Proposal |
|---|---|---|---|---|
| 3 | 2.927% | 0.214% | 0.966% | 0.123% |
| 5 | 1.895% | 0.165% | 0.441% | 0.055% |
| 7 | 0.770% | 0.092% | 0.232% | 0.043% |
| 9 | 0.111% | 0.023% | 0.092% | 0.018% |
| 11 | 0.073% | 0.056% | 0.083% | 0.009% |
| 13 | 0.028% | 0.086% | 0.018% | 0.009% |

**Table 3** The comparison with respect to the size of support set.

| Size | DLN [12] | LSTM [13] | Attention [14] | Proposal |
|---|---|---|---|---|
| 100 | 3.785% | 0.267% | 0.198% | 0.169% |
| 200 | 1.181% | 0.091% | 0.087% | 0.078% |
| 300 | 0.091% | 0.084% | 0.069% | 0.036% |
| 400 | 0.028% | 0.034% | 0.018% | 0.008% |
| 500 | 0.067% | 0.023% | 0.041% | 0.009% |

## VI. Discussion

The aforementioned results show that the size of support set, the extraction depth, and the embedding dimension are significantly important to overcome the generalization issue of the learning model. Although some partial features are unimportant for describing behaviors, detecting attacks and classifying traffics, it is crucial to imbed bytes, extract packet, and preprocess traffic for getting the relationship to be cleaned up, labelled, preprocessed, extracted without the loss of spatial and temporal locality with consideration of the location invariance, compositionality and statistic measurement. To face the challenges of scanning, sampling, formatting, padding and classifying anomaly multimedia traffics, the resources from mobile devices and tablets can be federated as clients that have their own local training data. The local dataset on each particular client is typically non-independent and identically distributed (non-IID), because the massively distributed users are personalized. The parameters of a shared model are exchanged via the federation of servers and clients. It reduced the risks of privacy for different clients. In a nutshell, the experiments demonstrate the proposal is effective and efficient to get the information of non-IID private data in each client. As one of implications of this research, it has potential to be extended to a general methodology on network structure learning for distributed media service systems and in-network computing systems.

## VII. Conclusion

A novel algorithm for multiparty privacy learning in graynet was proposed by exploring and testing the principles of meta-generalization. An action-relational graph was formulated as a potential tool to get better understanding of graynet. The learning algorithm was implemented to improve the performance of anomaly multimedia traffic identification. This letter contributed an approach of on-demand multiparty computation to deal with non-IID data. This letter discovered the inherent attributes of the multiparty privacy learning model in graynet to reduce the generalization error. Though the evaluation is limited to the anomaly multimedia traffic identification, it has potential to be extended to more research fields.

For the future work, it is meaningful to study the secure scalability on hierarchical, neighborhood and federated structure learning with regard to deterministic latencies of multimedia service, dynamic geographical locations and inter-domain transfer on machine learning parameters.

## Appendix

The input vector of multiparty privacy learning in graynet is

$$X_j = (x_1, x_2, ..., x_{d_j})^T.$$

where $d_j$ is the size of support set of the $j$th neuron. The output scalar is

$$y_j = f\left(C_j X_j + b_j\right).$$

where $f$ is the activation function, $C_j$ is a vector for connection, and $b_j$ is a scalar for bias. The negative input vector is

$$Y_j = (y_1, y_2, ..., y_{d_j})^T.$$

where $d_j \leq d_{\max}$, and $d_{\max}$ is a constant of the maximum degree. The negative output scalar is

$$x_j = g\left(C_j Y_j + b_j\right).$$

where $g$ is the activation function in the negative direction. The error function with its minimization is defined by

$$E(\Theta) = \sum_j \left\| y - y_j \right\|^2 + \sum_j \lambda_j \left\| x - x_j \right\|^2,$$

$$E^* = \min_{\Theta} E(\Theta),$$

\* satoshint@yeah.com 

$$\Theta^* = \arg\min_\Theta E(\Theta).$$

where $x$ is a sample from private and public datasets, $y$ is a known label as correct identification, $E(\Theta)$ is a user-defined error function, $\lambda_j$ is the coefficient for unlabeled training error, $\Theta^*$ is an optimized set of parameters, and $E^*$ is an optimal error function based on decentralized data [19-20].